\documentclass[conference]{IEEEtran}
\IEEEoverridecommandlockouts
\usepackage{cite}
\usepackage{amsmath,amssymb,amsfonts}
\usepackage{algorithmic}
\usepackage{graphicx}
\usepackage{textcomp}
\usepackage{xcolor}
\usepackage{booktabs}
\usepackage{subcaption}
\usepackage{url}
\usepackage[final]{changes}
\def\BibTeX{{\rm B\kern-.05em{\sc i\kern-.025em b}\kern-.08em
    T\kern-.1667em\lower.7ex\hbox{E}\kern-.125emX}}

\begin{document}

\title{Empirical evaluation of Time Series
Foundation Models for Day-ahead and
Imbalance Electricity Price Forecasting in
Belgium%
\thanks{Hussain Kazmi acknowledges support from the CELSA PREDDICT project. Margarida Mascarenhas acknowledges support from 1S39625N.}}

\author{
\IEEEauthorblockN{Chi Bui}
\IEEEauthorblockA{\textit{ELECTA-ESAT (KU Leuven)} \\
% \textit{KU Leuven}\\
Leuven, Belgium \\
thimaichi.bui@student.kuleuven.be}
\and
\IEEEauthorblockN{Maria Margarida Mascarenhas}
\IEEEauthorblockA{\textit{ELECTA-ESAT (KU Leuven)} \\
Leuven, Belgium \\
margarida.mascarenhas@kuleuven.be}
\\
\IEEEauthorblockN{Hussain Kazmi}
\IEEEauthorblockA{\textit{ELECTA-ESAT (KU Leuven)} \\
Leuven, Belgium \\
hussain.kazmi@kuleuven.be}
\and
\IEEEauthorblockN{Arnaud Verstraeten}
\IEEEauthorblockA{\textit{ELECTA-ESAT(KU Leuven) and Gridual}\\
Leuven, Belgium \\
arnaud.verstraeten@kuleuven.be}
}

\maketitle

\begin{abstract}
Recent advances in Time Series Foundation Models (TSFMs) promise zero-shot forecasting capabilities with minimal task-specific training. While these models have shown strong performance across generic benchmarks, their applicability in volatile, complex electricity markets remains underexplored. Addressing this gap, this study provides a systematic empirical evaluation of several TSFMs, specifically Chronos-2 and Chronos-Bolt (developed by Amazon), and TimesFM 2.5 (provided by Google), for forecasting Belgian day-ahead and imbalance electricity prices. For both considered markets, Chronos-2 in ARX mode produces the most accurate forecasts. \deleted{However,}\replaced{Compared with the best ensemble prediction from other machine learning methods, Chronos-2's Mean Absolute Error (MAE) is 5\% lower for the day-ahead market. In contrast, the model yields 10\% higher MAE predicting imbalance prices across all forecast horizons, except for the two-hour-ahead horizon}{15.6\% higher. Similarly, Chronos-2 yields an}. Moreover, we find that TSFMs exhibit genuine zero-shot forecasting skills but still struggle under extreme market conditions.

\end{abstract}


\section{Introduction}

In Belgium, more than a quarter of national electricity consumption is traded on the day-ahead market (DAM), where prices are set for the next day's delivery. The imbalance market (IMB) penalizes deviations from the DAM nominations. Historically, imbalance prices have exhibited higher volatility than DAM prices due to market design and trading time granularity, which creates higher opportunities for trading and arbitrage. These differences introduce distinct forecasting challenges for each market.

Complex forecasting models, such as ensembles of deep neural networks and regularized linear models, have shown cutting-edge performance in both day-ahead and imbalance markets \cite{Lago2021}, but they often entail high computational costs and require substantial engineering for maintenance, especially when faced with different bidding zones and non-stationarities. Furthermore, while day-ahead price forecasting is a well-established field, imbalance price forecasting is not yet at the same stage of development.

Time Series Foundation Models (TSFMs), often developed using transformer-based architectures, offer a universal forecasting framework alternative with minimal tuning requirements. In fact, TSFMs claim to leverage cross-domain pretraining for zero-shot forecasting, reduce model training effort, and improve adaptability. As such, they provide an interesting alternative to building specialised forecast models for each individual market and bidding zone. However, in practice, there is still a lack of systematic evaluation of whether such TSFMs can be used in the dynamic conditions seen in different electricity markets, especially under extreme volatility. 

This study evaluates the performance of such TSFMs for the Belgian DAM and IMB markets, benchmarking against state-of-the-art ensemble-based baselines.

\section{Forecasting in electricity markets}

\subsection{Day-ahead market}

The day-ahead market is the primary short-term electricity market in Europe. In this market, participants submit bids and offers for electricity delivery for each hour of the following day. The market is cleared through a price auction in which supply and demand curves are aggregated across participating bidding zones. The resulting market-clearing price determines the electricity price for each delivery hour of the next day. 

Most European countries, including Belgium, participate in the Single Day-Ahead Coupling (SDAC) operated through the EPEX SPOT exchange. The gate-closure time for the day-ahead market is 12:00 p.m. CET. Prior to this deadline, market participants must forecast electricity prices in order to support bidding strategies, portfolio optimization, and risk management. Day-ahead electricity price forecasting has been extensively studied. Comprehensive overviews of electricity price forecasting methods and evaluation practices are provided by Weron \cite{WERON20141030} and, more specifically, for day-ahead markets, by Lago \textit{et al.} \cite{LAGO2021116983}, \cite{LAGO2018386}.

\subsection{Imbalance market}
After the closure of day-ahead and intraday markets, balance responsible parties (BRPs) should ensure that their portfolio balance is in accordance with nominated volumes. Any deviations from the planned schedule are registered by the transmission system operator (TSO), and they are penalized or remunerated based on the imbalance price. In most European countries (including Belgium), the imbalance settlement period is equal to 15 minutes. As the sign of the collective system imbalance determines the need for upward or downward regulation volumes, the imbalance price can fluctuate strongly depending on the balancing state of the system. In addition, the system imbalance can be difficult to predict due to the uncertain output of variable renewables, which in turn leads to volatile imbalance prices. In the literature, there are two main approaches to forecasting imbalance prices. The first approach is to use a fundamental forecasting method based on the balancing regime, such as in \cite{bunn, dumas}. Another forecasting approach is to use data-driven machine learning models such as gradient boosting or neural networks \cite{ganesh_bunn, bottieau, deng}.

\subsection{TSFMs in electricity markets}
According to the technical reports introducing Amazon's Chronos \cite{ansari2024chronos} and Google's TimesFM \cite{Das2024}, such TSFMs have demonstrated strong zero-shot forecasting performance across a wide range of public benchmark datasets. These evaluations include datasets from multiple domains, such as retail demand, financial time series, and energy-related datasets, including Australian Electricity Demand, ETT (Electricity Transformer Temperature), and ERCOT Load. A recent study conducted by Timothée et al. \cite{hornek2025benchmarking} investigated the forecasting performance of TSFMs for DAM across multiple European bidding zones. Although this work provides valuable insights into the potential of foundation models for the electricity price forecasting task, the analysis focuses primarily on day-ahead markets.
In contrast, imbalance markets have received considerably less attention in the forecasting literature despite their importance for real-time system balancing and trading strategies. To our best knowledge, no existing study has systematically evaluated the performance of TSFMs for imbalance price forecasting.
This work aims to bridge these research gaps by providing an empirical evaluation of TSFMs for both the Belgian DAM and IMB market. By comparing zero-shot forecasts from prominent TSFMs with state-of-the-art baselines, we provide insights into the practical applicability of TSFMs in electricity market forecasting.

\section{Methodology}
This section describes the forecasting framework used to evaluate the performance of TSFMs relative to traditional forecast approaches. For each market, the models produce point forecasts for fixed horizons that align with the market's operational requirements. In the day-ahead market, the forecasting horizon consists of the 24 hourly prices of the next day. In the imbalance market, forecasts are generated for the next eight 15-minute settlement periods, corresponding to a two-hour horizon.

\subsection{Baseline models}

For the day-ahead market, a weekly persistence model is used. The forecasted price for each hour of the next day is equal to the observed price from the same hour one week earlier. For the imbalance market, two different naive models are used. The persistence model predicts the most recent imbalance price for the full forecast horizon. The second naive model uses the day-ahead price as the imbalance price forecast.

\subsection{Baseline data-driven models}
In addition to naive benchmarks, two state-of-the-art data-driven forecasting models are considered: the Lasso-enhanced AutoRegressive (LEAR) model and a Deep Neural Network (DNN) model. The LEAR model is a linear autoregressive model with Lasso regularization. Due to its linear structure, the LEAR model is computationally efficient and offers a higher degree of interpretability compared with the DNN model, which is a nonlinear alternative capable of learning more complex relationships between input variables and electricity prices. 

Both models are trained using a rolling calibration window of one year and are periodically recalibrated to account for evolving market dynamics. Furthermore, both models incorporate feature selection mechanisms. The LEAR model performs feature selection implicitly through regularization, while the DNN model includes feature selection as part of the hyperparameter optimization process. Prior to inference, the DNN model undergoes hyperparameter optimization over the search space presented in Table \ref{tab:dnn_hyperparameters}. More detailed descriptions of these models and their forecasting performance in electricity markets can be found in \cite{MASCARENHAS2026127077, LAGO2021116983}.
\added{In addition, model forecast ensembles are used to construct baseline results, as ensembling is a well-established way to improve forecasting performance \cite{WERON20141030}. Here, we simply average the forecasts from individual models to produce the ensemble forecast.
}

\subsection{TSFMs}
TSFMs are large models with transformer architectures and millions of parameters, similar to large language models (LLMs) but for the time-series domain. These models are claimed to offer a universal forecasting framework with minimal tuning requirements. TSFMs can be categorized into Pre-trained and LLM-based models \cite{TSFMtutorialsurvey}, \cite{TSFM-Bench}. While LLM-based models aim to adapt existing large language model architectures for time series forecasting tasks, pre-trained models are trained on diverse, multi-domain time series datasets. This study's scope is restricted to pre-trained models. 

Pre-trained TSFMs utilize various techniques to transform time points into \textit{tokens} that form input layers, which are then trained as LLMs. Moreover, in zero-shot inference, \textit{context length} \replaced{controls how far back into the past values that foundation models have accessed to and perform attention mechanism, and can be considered to be equivalent to maximum lag order in traditional time series analysis}{is used interchangeably with lag in traditional time series analysis}. We use two inference modes: AR (univariate, using only historical target data) and ARX (multivariate, incorporating cointegrated time series) from two libraries \texttt{chronos-forecasting} and \texttt{timesfm}. The ARX mode is implemented only for the Chronos-2 model.

\paragraph{Chronos (Amazon)} Chronos was first introduced in 2024 with two main variants, i.e, Chronos (original variant) and Chronos-bolt (fast inference variant), each having its own derivatives based on model size. In this study, we chose to experiment with two representative models for each variant,  Chronos-2 and Chronos-bolt (base), having 120M and 205M parameters, respectively. Standard encoder-decoder architecture is employed for building Chronos-bolt while Chronos-2 adopts encoder-only design.

A key innovation in Chronos is the transformation of continuous time series into discrete token sequences by applying quantization to map numerical values to discrete bins defined by learned or fixed scaling rules \cite{ansari2024chronos},\cite{ansari2025chronos2univariateuniversalforecasting}. This discretization enables the model to treat time-series forecasting as a language modeling problem, predicting the next token given previous tokens. After inference, tokens are mapped back to numerical values through inverse scaling.

\paragraph{TimesFM (Google)}
Google's TimesFM is among the pioneering models in this emerging research of TSFMs, with TimesFM 1.0 introduced in early 2024. In this study, we use the latest release, TimesFM 2.5 (2025), which contains approximately 200 million parameters and offers improved documentation and ease of implementation. Unlike Chronos, TimesFM adopts decoder-only architecture and forms tokens by grouping temporally adjacent time steps regardless of their numerical magnitude.

\subsection{Input features}
Table \ref{tab:variables_used} contains the input features used in the forecasting models. For day-ahead price forecasting, the input covariates include lagged electricity prices (autoregressive component) from the previous two days and the previous week, lagged exogenous variables from the previous day and the previous week, as well as for the day being forecasted, and calendar features (a holiday dummy variable and dummy variables for the day of the week). 

In the case of imbalance price forecasting, the inputs include lagged imbalance prices and system imbalance values from the past hour, and past and future day-ahead prices. The lookahead for future input information is the same as the forecast horizon, which is equal to two hours.

\begin{table}[htbp]
    \centering
    \caption{Input features used in day-ahead and imbalance price forecasting}
    \begin{tabular}{lccc}
        \toprule
         \textbf{Data} & \textbf{Frequency} & \textbf{Unit} & \textbf{Forecast} \\
         \midrule
         DAM price & Hourly & EUR/MWh & DAM, IMB \\
         Wind generation forecast & Hourly & MW & DAM \\
         Day-ahead load forecast & Hourly & MW & DAM \\
         Solar generation forecast& Hourly & MW & DAM \\
         Humidity forecast& Hourly & \% & DAM \\
         Temperature forecast& Hourly & \textdegree C & DAM \\
         Imbalance price & 15 minutes & EUR/MWh & IMB \\
         System imbalance & 15 minutes & MW & IMB \\
         \bottomrule
    \end{tabular}
    
    \label{tab:variables_used}
\end{table}

\subsection{Evaluation metrics}
We use Mean Absolute Error (MAE) and Root Mean Squared Error (RMSE) to evaluate the accuracy of the forecasts, which are common metrics used in electricity price forecasting literature. \cite{WERON20141030},  \cite{LAGO2021116983}, \cite{LAGO2018386}, \cite{MASCARENHAS2026127077}. %todo: references

\begin{equation}
\text{MAE}=\frac{\sum_{t=1}^{\text{T}} \left |\hat{y}_t - y_t \right|}{\text{T}} \label{eq} \quad \text{(Unit: EUR/MWh)}
\end{equation}
\begin{equation}
\text{RMSE}=\sqrt{\frac{\sum_{t=1}^{\text{T}} \left (\hat{y}_t - y_t \right)^2}{\text{T}}} \label{eq} \quad \text{(Unit: EUR/MWh)}
\end{equation}

\noindent where $y_t$ is the true value and $\hat{y}_t$ is the predicted value.
Lower values of MAE and RMSE are desWe use Mean Absolute Error (MAE) and Root Mean Squared Error (RMSE) to evaluate the accuracy of the forecasts, which are common metrics used in electricity price forecasting literature. \cite{WERON20141030},  \cite{LAGO2021116983}, \cite{LAGO2018386}, \cite{MASCARENHAS2026127077}. %todo: references

\begin{equation}
\text{MAE}=\frac{\sum_{t=1}^{\text{T}} \left |\hat{y}_t - y_t \right|}{\text{T}} \label{eq} \quad \text{(Unit: EUR/MWh)}
\end{equation}
\begin{equation}
\text{RMSE}=\sqrt{\frac{\sum_{t=1}^{\text{T}} \left (\hat{y}_t - y_t \right)^2}{\text{T}}} \label{eq} \quad \text{(Unit: EUR/MWh)}
\end{equation}

\noindent where $y_t$ is the true value and $\hat{y}_t$ is the predicted value.
Lower values of MAE and RMSE are desirable as shows better approximate between predictions and true values. However, these metric are prone to outliers. irable as shows better approximate between predictions and true values. However, these metric are prone to outliers. 

\section{Experiments and Results}

\subsection{Case study setup}
For the Belgian day-ahead market, year 2023 is used for initial calibration and 2024 is used as the out-of-sample test year; during testing, the one-year calibration window is rolled forward day-by-day while keeping its length fixed. For the Belgian imbalance market, 2022 is used for initial calibration, and 2023 is used as the out-of-sample test year, with the same fixed-length rolling-window procedure and monthly recalibration. The time period under consideration has been cross-checked with the list of disclosed pretrained datasets that all tested TSFMs were trained on to avoid the well-known data leakage problem of TSFMs. A high-level summary of the time series used in the analysis is included in Table \ref{tab:variables_used}.
\added{Different context length configurations for TSFMs have been experimented with, and the best performance is reported, to be specific, 8192 hours of context (almost one year) are used for Chronos-2 and Chronos-bolt in DAM, and 2048 quarter-hours (approximately 1 month) for IMB forecast, while TimesFM uses a context length of 1024 hours in both markets.}
\subsection{Forecast accuracy}

\begin{table}[htbp]
    \centering
    \caption{Forecast MAE for naive persistent model, data - driven models, TSFMs and their ensembles in DAM and three horizons in IMB market}
    \begin{tabular}{lcccc}\toprule
         \textbf{Market} & \textbf{DAM} & \multicolumn{3}{c}{\textbf{IMB}} \\\cmidrule(lr){3-5}
         & & \textbf{QH1} & \textbf{QH4} & \textbf{QH8} \\
         \midrule
         Persistent model & 28.22 & 76.8 & 104.68 & 124.57 \\
         \midrule
         Chronos-bolt    & 16.03 & 75.42 & 92.44 & 97.8 \\
         Chronos 2 - AR  & 15.61 & 72.91 & 90.3 & 95.43 \\
         Chronos 2 - ARX & \replaced{12.43}{ 15.28} & \replaced{70.1}{ 71.33} & \replaced{87.73}{ 88} & \replaced{93.72}{ 93.73} \\
         TimesFM 2.5     & 16.95 & 74.39 & 92.15 & 97.78 \\
         TSFMs Ensemble  & \replaced{14.43}{15.49} & \replaced{72.12}{ 72.46} & \replaced{89.58}{  89.59} & \replaced{95.13}{ 95.01} \\
         \midrule
         LEAR            & 13.74 & 68.74 & 88.01 & 94.71 \\
         DNN             & 14.39 & 68.69 & 87.87 & 95.13 \\
         ML Ensemble        &  13.22 & 67.19 & 87.3 & 93.96 \\
         \midrule
         Chronos 2 \& ML      & \replaced{\underline{12.3}}{\underline{ 12.81}}   & \replaced{\underline{67.09}}{\underline{ 66.76}}  &\replaced{\underline{86.45}}{\underline{86.55}}  & \replaced{\underline{92.83}}{\underline{ 93}}\\
         \bottomrule
    \end{tabular}
    \label{tab:mae_acc}
\end{table}
\begin{table}[htbp]
    \centering
    \caption{Forecast RMSE for naive persistent model, data - driven models, TSFMs and their ensembles in DAM and three horizons IMB market}
    \begin{tabular}{lcccc}\toprule
         \textbf{Model} & \textbf{DAM} & \multicolumn{3}{c}{\textbf{IMB}} \\\cmidrule(lr){3-5}
         & & \textbf{QH1} & \textbf{QH4} & \textbf{QH8} \\
         \midrule
         Persistent model & 40.99 & 130.81 & 162.45 & 183.96 \\
         \midrule
         Chronos-bolt    & 24 & 114.21 & 132.48 & 138.27 \\
         Chronos 2 - AR  & 23.27 & 115.08 & 132.3 & 137.52 \\
         Chronos 2 - ARX & \replaced{19.36}{ 22.89} & \replaced{110.77}{ 110.69} & \replaced{130.35}{ 129.53} & \replaced{136.61}{ 136.09} \\
         TimesFM 2.5     & 25.07 & 113.86 & 133.55 & 139.68 \\
         TSFMs Ensemble  & \replaced{21.74}{23.04} &  \replaced{111.17}{111.91} & \replaced{130.2}{130.35} & \replaced{136.18}{136.15}  \\
         \midrule
         LEAR            & 20.38 & 105.58 & 126.26 & 133.36 \\
         DNN             & 20.98 & 108.32 & 128.87 & 136.61 \\
         ML Ensemble        &  19.72  & \underline{104.15} & 126.19 & 133.52 \\
         \midrule
         Chronos 2 \& ML  &  \replaced{\underline{18.91}}{\underline{ 19.58}}  &104.3 & \replaced{\underline{126.05}}{\underline{125.86}} & \replaced{\underline{132.98}}{\underline{133.02}} \\
         \bottomrule
    \end{tabular}
    \label{tab:rmse_acc}
\end{table}

\begin{figure}[htbp]
    \centerline{\includegraphics[scale=0.3]{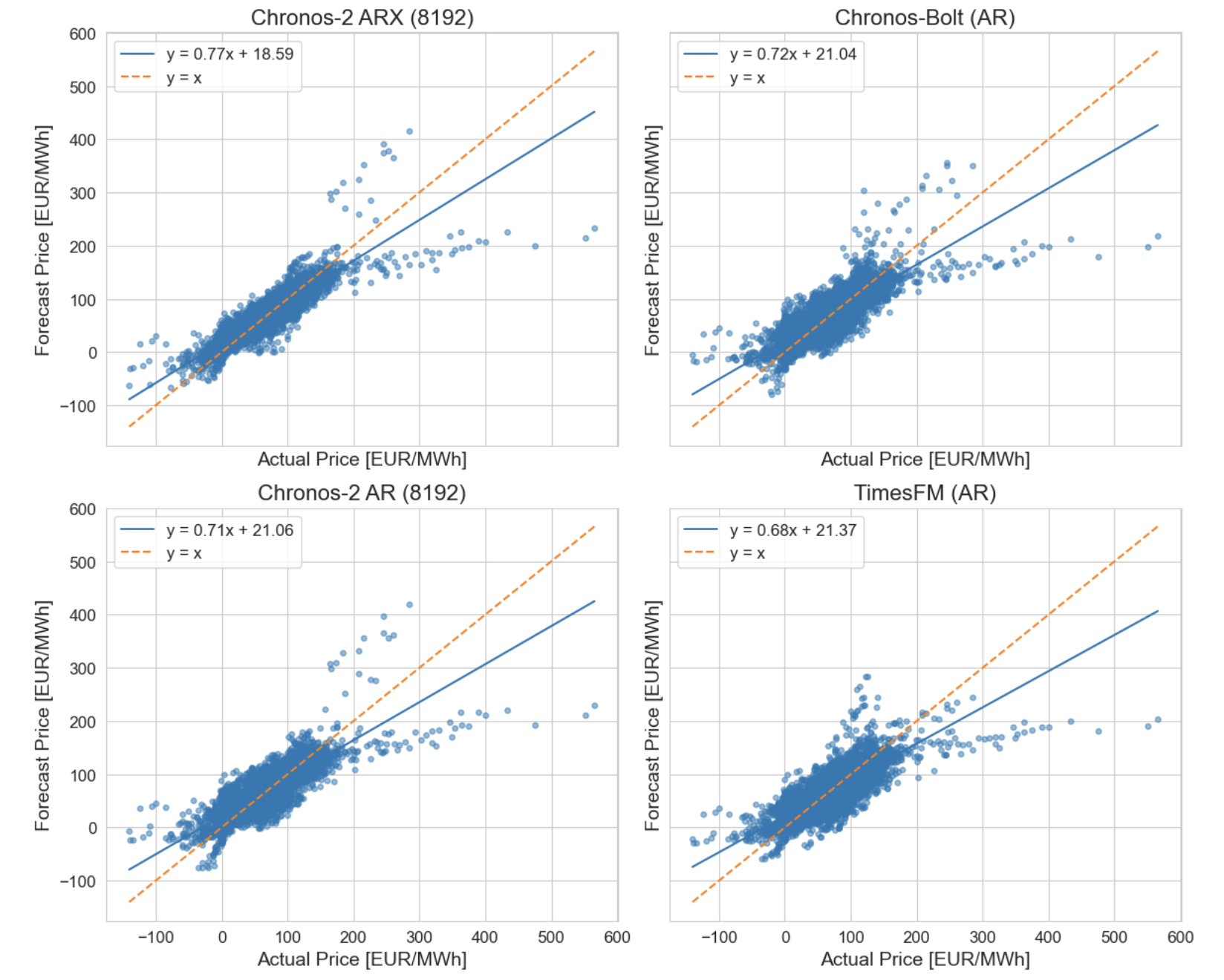}}
   
    \caption{DAM: TSFMs forecasts vs. Actual prices}
    \label{fig:dam_corr_plot}
\end{figure}

\begin{figure}[htbp]
    \centerline{\includegraphics[scale=0.3]{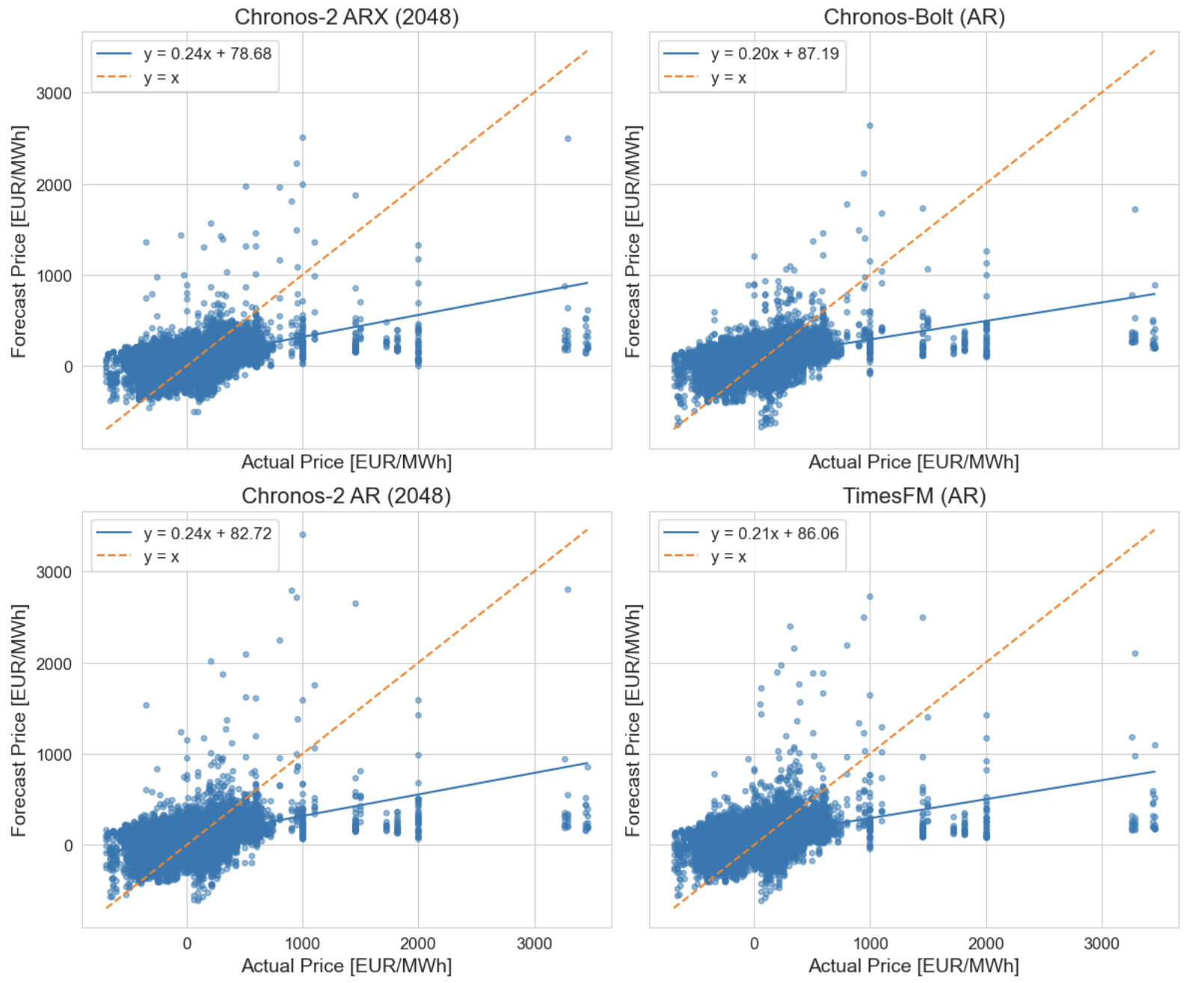}}
   
    \caption{IMB: TSFMs forecasts vs. Actual prices}
    \label{fig:imb_corr_plot}
\end{figure}

The MAE and RMSE for day-ahead price forecasts and selected quarter-hour imbalance price forecasts are summarized in Tables \ref{tab:mae_acc} and \ref{tab:rmse_acc}. \added{These two metrics are expressed in EUR/MWh, enabling direct comparison to the average DAM price (73.38 EUR/MWh in 2024) and IMB price (96.72 EUR/MWh in 2023).} 
Overall, among TSFMs, Chronos-2 in ARX mode produced the most accurate forecasts for both DAM and IMB prices. It is also worth noting that combining forecasts from all TSFMs yields worse results than individual Chronos-2 (ARX mode) forecasts for both markets, due to high correlation among TSFMs forecasts seen in Figures \ref{fig:dam_corr_plot} and \ref{fig:imb_corr_plot}.

Regarding DAM, when using Chronos-2 with covariates, errors are reduced\replaced{  significantly, in particular, 20\% smaller than}{ slightly} AR mode, \replaced{and even slightly improved in relative to}{ but remain approximately 10-15\% worse than} trained and calibrated LEAR and DNN models with exactly the same set of features\added{, indicating the acclaimed outstanding capabilities of the model on leveraging covariates inputs}. \deleted{Only w}When combining forecasts from Chronos-2 and baseline data-driven models (LEAR and DNN) were we able to achieve negligibly lower MAE and RMSE of 12.29 EUR/MWh and 18.91 EUR/MWh, respectively. However, conducting the Diebold-Mariano test between the Chronos-2 \& ML ensemble, and the ML ensemble reveals that the improvement is not significant.

\begin{figure}[htbp]
\centerline{\includegraphics[scale=0.45]{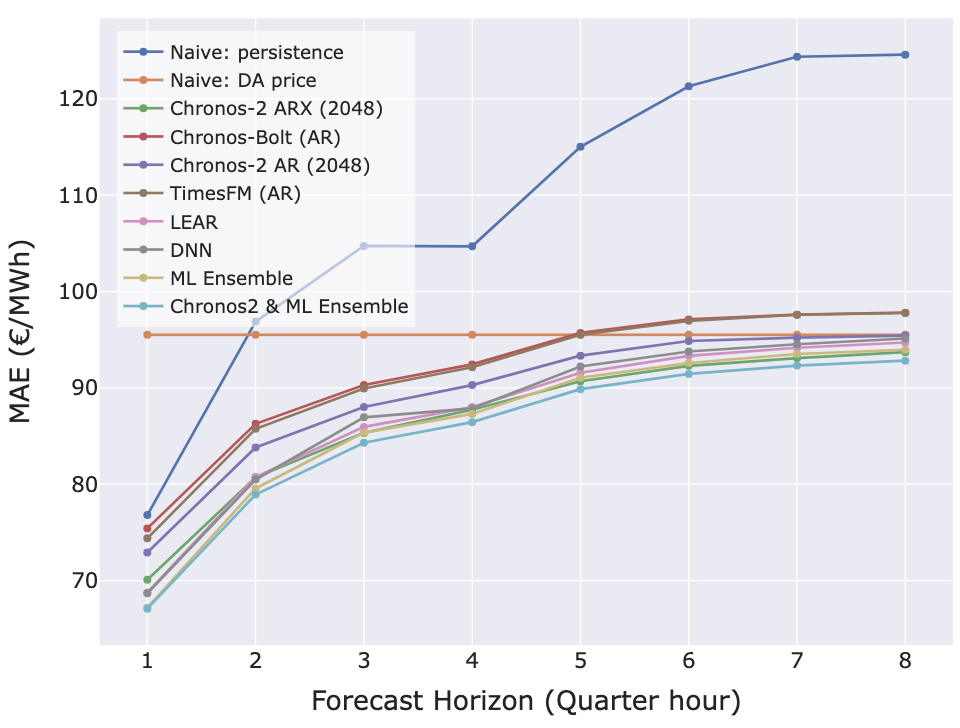}}
\caption{IMB: MAE of different models across all forecast horizons}
\label{fig:imb_mae}
\end{figure}

Regarding IMB, as shown in Figure \ref{fig:imb_mae}, TSFMs outperformed the naive persistent model across all quarter hours. Except for Chronos-bolt and TimesFM for horizons from 5 to 8 quarter-hour, other TSFMs setups also surpass naive model of predicting Day-ahead price across all forecast horizon. However, when being benchmarked against trained data-driven models, only Chronos-2 in ARX mode outperformed at longer forecast horizons, i.e., the eighth quarter hour (as per Table \ref{tab:mae_acc} and \ref{tab:rmse_acc}). It is also clear that when combining forecasts from Chronos-2 in ARX mode with ML baseline models, the ensemble yields slightly more accurate forecasts than the ML ensemble alone, as evidenced by lower MAE and RMSE across all forecast horizons.

\subsection{Errors Analysis}
\paragraph{DAM}

To analyze how each model performs across different market conditions, we report its accuracy for 1, 5, and 10 DAM price deciles in Table \ref{tab:decile_acc}. \deleted{Although TSFMs evaluated individually cannot outperform data-driven baseline models, we still see improvements when incorporating Chronos-2's forecasts with ML models.} \added{Alhough Chronos-2 in ARX mode outperformed ensemble forecasts of data-driven models in terms of accuracy evaluated for the test year, the model struggles in unfavorable occasions - 1st and 10th decile}. The Chronos-2 \& ML combination consistently yields lower MAE and RMSE across all price deciles compared to individual models, indicating their potential use of Chronos-2 in combination with currently operating models in predicting DAM prices under extreme market conditions, eventually benefiting risk-aware strategic planning.

\begin{table}[htbp]
    \centering
    \caption{Accuracy across price deciles in DAM}
    \begin{tabular}{lrrr|rrr}
\toprule

\textbf{Metrics} & \multicolumn{3}{c}{\textbf{MAE}} & \multicolumn{3}{c}{\textbf{RMSE}} \\
\cmidrule(lr){2-4} \cmidrule(lr){5-7}
\cmidrule(lr){2-7}
\textbf{Model/ Decile}& \textbf{1} & \textbf{5} & \textbf{10} & \textbf{1} & \textbf{5} & \textbf{10} \\
\midrule

Chronos-bolt & 24.16 & 12.34 & 26.51 & 32.36 & 16.59 & 44.07 \\
Chronos 2 - AR & 23.42 & 12.03 & 26.11 & 31.43 & 16.23 & 42.87 \\

Chronos 2 - ARX & 19.36 & \underline{9.08} & 21.28 & 25.29 & 12.01 & 39.06 \\
TimesFM 2.5 & 24.27 & 12.96 & 28.24 & 31.34 & 18.01 & 45.41 \\

TSFMs Ensemble & 21.93 & 10.97 & 24.60 & 28.97 & 14.72 & 41.31\\
\midrule
LEAR & 19.78 & 10.78 & 22.25 & 25.71 & 14.13 & 39.42 \\
DNN & 20.01 & 10.87 & 23.15 & 25.99 & 14.58 & 39.19 \\
ML Ensemble&19.17 & 10.03 & 21.8 & 24.81 & 13.37 & 38.24\\
\midrule

Chronos-2 \& ML &\underline{18.84} & 9.15 & \underline{20.86} &\underline{24.4} & \underline{12} & \underline{38}\\
\bottomrule
\end{tabular}
    \label{tab:decile_acc}
\end{table}

\paragraph{IMB}
Similar to what has been done with IMB, we also segregate imbalance price into deciles and analyze the MAE of each model's forecasts for the price's 10th and 90th percentiles with forecast horizons of 1, 4, and 8 quarter hours (Table \ref{tab:decile_acc_imb}). TSFMs clearly struggle under extreme conditions, among which only Chronos-2 outperformed all models for the 8th quarter-hour forecast.

\begin{table}[htbp]
    \centering
    \caption{MAE for IMB forecasts under extreme conditions}
    \begin{tabular}{lrrr|rrr}
\toprule

 \textbf{Model}& \multicolumn{3}{c}{\textbf{1st Decile}} & \multicolumn{3}{c}{\textbf{10th Decile}} \\
\cmidrule(lr){2-4} \cmidrule(lr){5-7}
\cmidrule(lr){2-7}
  & \textbf{QH1} & \textbf{QH4} & \textbf{QH8} & \textbf{QH1} & \textbf{QH4} & \textbf{QH8} \\
\midrule

Chronos-bolt & 173.8 & 228.9 & 250.5 & 108.9 & 161.1 & 176.9 \\
Chronos 2 - AR & 162.5 & 220.9 & 240.5 & 100.6 & 153.6 & 174.2 \\
Chronos 2 - ARX & 164.8 & 213.9 & 235.3 & 101.7 & 155.4 & 177.5\\
TimesFM 2.5 & 171.1 & 229.5 & 250.5 & 106.4 & 160.5 & 178.4 \\
TSFMs Ensemble & 167.1 & 222.7 & 244.04 & 103.6 & 157.2 & 176.5\\
\midrule
LEAR & \underline{155.1} & \underline{205.9} & \underline{232.9} & 81.0 & 152.4 & 181.8 \\
DNN & 167.5 & 211.5 & 237.4 & \underline{78.7} & \underline{146.4} & \underline{173.9} \\
ML Ensemble&160.8 & 208.6 & 234.0 & 81.2 & 152.1 & 179.5\\
\midrule
Chronos-2 \& ML & 161.9 & 210.7 & 234.3 & 88.5 & 153 & 178.1 \\
\bottomrule
\end{tabular}
    \label{tab:decile_acc_imb}
\end{table}

\section{Conclusion}
This study presents an empirical evaluation of the zero-shot forecasting capabilities of Time Series Foundation Models (TSFMs) in the Belgian day-ahead (DAM) and imbalance (IMB) electricity markets. \deleted{The results show that while TSFMs achieve competitive performance relative to naive baseline models, data-driven models and their ensembles consistently outperform TSFMs under both normal and extreme price conditions. }\added{The results show that, by operating in ARX mode, Chronos-2 was able to utilize covariates and even outperformed both naive and data-driven baselines, as well as their ensemble forecasts, for DAM. Moving forward, the underlying reasons for this impressive improvement in predictive accuracy pose an intriguing question, where an in-depth dive into model architecture and pre-trained datasets is necessary to inspect how covariate information is used and to avoid any data leakage problems.}
\replaced{Furthermore}{Nevertheless}, combining TSFMs' forecasts with data-driven models improves predictive power for both DAM and IMB prices over a one-year test period, suggesting that TSFMs can serve as a valuable complementary component in ensemble forecasting frameworks, even though further consideration is needed for handling extreme price markets. In addition to forecasting accuracy, TSFMs offer practical advantages in terms of scalability and operational efficiency, as they require minimal feature engineering and no model retraining. These characteristics make them promising tools for rapid deployment across multiple electricity markets and bidding zones.

\bibliography{references}
\newpage
\appendix

\begin{table}[ht]
\centering
\footnotesize
\caption{Hyperparameter search space for the DNN model.}
\label{tab:dnn_hyperparameters}
\begin{tabular}{ll}
\hline
\textbf{Parameter} & \textbf{Search space} \\
\hline
Neurons (layer 1) & $[50, 500]$ \\
Neurons (layer 2) & $[25, 400]$ \\
Dropout & $[0,1]$ (uniform) \\
Batch normalization & \{True, False\} \\
Regularization & \{None, L1\} \\
L1 penalty $\lambda$ & $[10^{-5}, 1]$ (log-uniform) \\
Learning rate & $[5\times10^{-4}, 0.1]$ (log-uniform) \\
Random seed & $[1,1000]$ \\
Activation function & \{ReLU, Softplus, Tanh, SELU, \\
& LeakyReLU, PReLU, Sigmoid\} \\
Weight initialization & \{Orthogonal, Lecun-uniform, \\
& Glorot-uniform, Glorot-normal, \\
& He-uniform, He-normal\} \\
Scale $X$ & \{None, Norm, Norm1\} \\
Scale $Y$ & \{None, Norm, Norm1\} \\
Day-of-week dummy & \{True, False\} \\
Holiday dummy & \{True, False\} \\
Price lag $D-1$ & \{True, False\} \\
Price lag $D-2$ & \{True, False\} \\
Price lag $D-3$ & \{True, False\} \\
Price lag $D-7$ & \{True, False\} \\
Exogenous variable $i$ ($D$) & \{True, False\} \\
Exogenous variable $i$ ($D-1$) & \{True, False\} \\
Exogenous variable $i$ ($D-7$) & \{True, False\} \\
\hline
\end{tabular}
\end{table}

\vspace{12pt}
\color{red}

\end{document}